\documentclass[11pt]{article}

\usepackage{amsmath, amssymb, amsfonts, mathtools, esint, stmaryrd}

\usepackage{bm}             % 粗体数学符号
\usepackage{bbold}          % 黑板字体
\usepackage{textcomp}       % 文本符号
\usepackage{mathrsfs}       % 花体数学字体

\usepackage{graphicx, overpic, float, epsf}  % 图像支持
\usepackage{subcaption, caption}             % 子图与标题
\usepackage{tikz}                             % 绘图
\usepackage{multirow, diagbox}               % 表格支持

\usepackage[colorlinks=true]{hyperref}
\hypersetup{pageanchor=false}
\usepackage[numbers,square,comma,sort&compress,merge]{natbib}  % 文献引用

\usepackage{appendix}
\usepackage{indentfirst}                     % 首段缩进

\usepackage[utf8]{inputenc}
\usepackage[margin=1in,letterpaper]{geometry}
\usepackage[affil-it]{authblk}               % 作者/机构格式

\usepackage{color}
\usepackage{comment}
\usepackage{ulem}                            % 注意：可能影响 \emph
\usepackage{url}
\usepackage{xfrac}

\usepackage{titlesec}
\usepackage{etoolbox}  % 提供 \ifappendix 判断

\newif\ifappendix
\appendixfalse  % 默认不是附录

\titleformat{\section}
  {\normalfont\Large\bfseries}
  {\ifappendix Appendix \thesection:\else \thesection\fi}
  {1em}
  {}

\numberwithin{equation}{section}
\let\originalleft\left
\let\originalright\right
\renewcommand{\left}{\mathopen{}\mathclose\bgroup\originalleft}
\renewcommand{\right}{\aftergroup\egroup\originalright}
\def\bea{\begin{eqnarray}}
\def\eea{\end{eqnarray}}

\usepackage{tensor}
\usepackage{physics}

\newcolumntype{P}[1]{>{\Centering\hspace{0pt}}p{#1}}
\newcolumntype{Z}{>{\centering\arraybackslash}X} %Z单元格居中

\begin{document}
\title{\bf Probing the Penrose Process: Images of Split Hotspots and Their Observational Signatures}
	
\author{Zhixing Zhao$^{1,2}$, Zhong-Ying Fan$^3$, Xiaobao Wang$^{4,2}$, Minyong Guo$^{1,2\ast}$, Bin Chen$^{5,6}$}
\date{}
	
\maketitle
\vspace{-15mm}

\begin{center}
{\it
$^1$ School of physics and astronomy, Beijing Normal University,
Beijing 100875, P. R. China\\\vspace{4mm}

$^2$ Key Laboratory of Multiscale Spin Physics (Ministry of Education), Beijing Normal University, Beijing 100875, China\\\vspace{4mm}

$^3$ Department of Astrophysics, School of Physics and Electronic Engineering, Guangzhou University, Guangzhou 510006, P. R. China\\\vspace{4mm}

$^4$ School of Applied Science, Beijing Information Science and Technology University, \\Beijing 100192, P. R. China\\\vspace{4mm}

$^5$ Institute of Fundamental Physics and Quantum Technology, \\ \& School of Physical Science and Technology, Ningbo University, Ningbo, Zhejiang 315211, China\\\vspace{4mm}

$^6$ School of Physics, \&  Center for High Energy Physics, Peking University, \\
No.5 Yiheyuan Rd, Beijing 100871, P. R. China\\\vspace{2mm}
}
\end{center}

\vspace{8mm}

\begin{abstract}

While theoretically established for decades, the Penrose process—energy extraction from rotating black holes—still lacks clear observational evidence. A promising theoretical framework posits magnetic reconnection in the ergosphere as a trigger, causing a plasmoid to separate into an escaping positive-energy fragment and an infalling negative-energy one. In this work, we investigate the observational imprints of this scenario. We treat the energized plasmoid as a hotspot and calculate its light curves for a realistic plasma magnetization. In particular, we further compare with the scenario in which the plasmoid, after fragmentation, falls into the black hole with positive energy, while all other conditions remain unchanged. Our results reveal that the process of fragmentation generates distinct flares, whose characteristics depend heavily on whether the infalling fragment carries negative or positive energy. We propose that these differences serve as identifiable signatures of the Penrose process.

\end{abstract}

\vfill{\footnotesize $\ast$ Corresponding author: minyongguo@bnu.edu.cn}

\maketitle

\newpage
\baselineskip 18pt
\section{Introduction}\label{sec1}

The Penrose process, a seminal mechanism for extracting rotational energy from black holes, was proposed over five decades ago \cite{Penrose:1969pc,Penrose:1971uk}. Its profound implications have inspired a wide array of extensions, including implementations in Reissner–Nordström spacetime \cite{Denardo:1974qis}, the collisional Penrose process \cite{Schnittman:2018ccg}, the Bañados–Silk–West effect \cite{Banados:2009pr}, and mechanisms involving spinning particles \cite{Zhang:2018gpn,Guo:2016vbt,Zhang:2016btg}, superradiant scalar fields \cite{Press:1972zz,Pani:2012vp}, and the Blandford–Znajek process \cite{Blandford:1977ds}. Despite this rich theoretical heritage, direct observational evidence for the Penrose process remains elusive, and a concrete observational strategy is yet to be established.

The observational challenge stems from two primary factors. First, the conditions for the process are exceptionally stringent, requiring, for instance, particle pairs to separate with a relative velocity exceeding $0.5c$ \cite{Bardeen:1972fi,Wald:1974kya}. Second, the process occurs deep within the ergosphere, perilously close to the event horizon, a region beyond the resolving power of current telescopes. Recent theoretical advances, however, suggest a promising pathway forward. It is now recognized that magnetic fields and plasmas permeate the environments of astrophysical black holes, and magnetic reconnection is likely to occur within current sheets inside the ergosphere \cite{Koide:2008xr,Asenjo:2017gsv,Zeng:2025vjt,Wang:2025pqh}. This reconnection can release copious energy, producing pairs of plasmoids—one ejected outward and the other captured by the black hole. If the infalling plasmoid carries negative energy-at-infinity, the Penrose process is effectively activated. The work of \cite{Comisso:2020ykg} quantified the power extraction and plasma energization efficiency from a Kerr black hole via this magnetic reconnection mechanism, sparking numerous subsequent investigations \cite{Wei:2022jbi,Li:2023htz,Chen:2024ggq,Fan:2024fcy,Khodadi:2022dff,Shen:2024sdr,Jiang:2024vgn,Zeng:2025olq,Fan:2024rsa,Zhang:2024rvk}. Notably, \cite{Camilloni:2024tny} demonstrated that such a plasmoid-driven Penrose process is energetically feasible under realistic astrophysical conditions, bolstering the case for its occurrence.

Concurrently, observational capabilities have taken a leap forward. The Event Horizon Telescope Collaboration has released  horizon-scale images of the supermassive black holes in M87* and Sgr A* \cite{EventHorizonTelescope:2019dse,EventHorizonTelescope:2022wkp}, while the GRAVITY instrument has achieved astrometric precision of $\sim 30 \mu\text{as}$ during infrared flares \cite{GRAVITY:2018ofz,abuter2018detection}. The appearance of these flares may be originated from the hotspot in orbital motion occurring at approximately nine gravitational radii \cite{GRAVITY:2020lpa}, offering a new window into the dynamics and spacetime geometry near the event horizon. As technology progresses, future observations will undoubtedly resolve phenomena even closer to the black hole, potentially enabling the detection of luminous events within the ergosphere itself.

Given the high likelihood of the magnetic-reconnection-driven Penrose process occurring in the ergosphere, a critical next step is to identify its unique observational imprints. This motivates our present work. We propose that the light curves and flare signatures associated with the reconnection-driven ejection and infall of plasmoids can serve as a viable probe. Specifically, we model a parent plasmoid in circular orbit that fragments via magnetic reconnection: the negative-energy component plunges into the black hole while the positive-energy counterpart escapes. Employing a hotspot model with plasma magnetisation parameters from \cite{Comisso:2020ykg}, we compute the corresponding images and light curves. Specifically, we examined two representative cases. In the first case, with a black hole spin of $a = 0.94$ and an observer azimuth of $\phi_0 = \pi / 2$, the infall of the negative-energy plasmoid gives rise to a single flare. In the second case, corresponding to a higher spin of $a = 0.99$ and an observer azimuth of $\phi_0 = 0$, the negative-energy plasmoid produces no flare. To identify the distinctive signatures arising from the Penrose process, we further analyse the scenario in which the plasmoid, after fragmentation, falls into the black hole with positive energy, while all other conditions remain unchanged. Through comparison, we find that the infalling plasmoids with positive and negative energies exhibit distinct flare characteristics at different observational azimuths, thereby providing unique observational features for identifying the Penrose process.

The remainder of this paper is organized as follows. Section~\ref{sec2} reviews the magnetic-reconnection-driven Penrose process in Kerr spacetime \cite{Comisso:2020ykg}. Section~\ref{sec3} introduces our hotspot model and the numerical ray-tracing methodology. Our results for two representative azimuthal angles are presented and discussed in Sec.~\ref{sec4}. Finally, we summarize and provide an outlook in Sec.~\ref{sec5}. Throughout this work, we use units where $c = G = 1$ and adopt the metric signature $(-,+,+,+)$.

\section{Magnetic Reconnection-Driven Penrose Process in Kerr Spacetime}\label{sec2}

In this section, we review the mechanism of the magnetic-reconnection-driven Penrose process as described in \cite{Comisso:2020ykg} and present the physical quantities associated with orbital calculations.

The metric of the Kerr spacetime in the Boyer–Lindquist (BL) coordinates is given by:  
\begin{equation}
ds^2 = -\left(1 - \frac{2M r}{\Sigma} \right) dt^2 - \frac{4M a r \sin^2\theta}{\Sigma} dt d\phi + \frac{\Sigma}{\Delta} dr^2 + \Sigma d\theta^2 + \frac{A}{\Sigma} \sin^2\theta d\phi^2 \,.\label{kerrg}
\end{equation}  
where  
\begin{equation}
\Sigma = r^2 + a^2 \cos^2\theta \,, \quad \Delta = r^2 - 2M r + a^2 \,, \quad A = (r^2 + a^2)^2 - a^2 \Delta \sin^2\theta \,.
\end{equation}
The two parameters appearing in the metric, $M$ and $a$, represent the black hole's mass and its spin parameter, respectively, with $|a| \leq M$. The location of the event horizon is determined by the roots of $\Delta = 0$, which are
\begin{equation}
r_{\pm} = M \pm \sqrt{M^2 - a^2} \,.
\end{equation}
Here, $r_{+}$ is the outer event horizon, which serves as the true boundary of the black hole. $r_{-}$ is the inner event horizon, also known as the Cauchy horizon. Moreover, the static limit surface in the Kerr spacetime is determined by the condition $g_{tt} = 0$, yielding  
\[
r_{\text{s}} = M + \sqrt{M^2 - a^2 \cos^2 \theta} \,.
\]
Given that $\cos^2 \theta \leq 1$, it follows that  
\[
r_{\text{s}} \geq r_+ \,.
\]
This implies that the static limit surface always lies outside the event horizon. The separation between the two surfaces attains its maximum at the equatorial plane ($\theta = \pi/2$) and gradually diminishes toward the poles ($\theta = 0$), where they coincide. The region between $r_+$ and $r_s$ is known as the ergosphere. Within the ergosphere, due to the condition $g_{tt} > 0$, negative energy orbits can exist, enabling the occurrence of the Penrose process, through which energy can be extracted from the black hole. Although the Penrose process theoretically demonstrates the possibility of extracting rotational energy from a black hole, its stringent requirements—such as the necessity for newly created particles to separate at a relative velocity exceeding $0.5c$ \cite{Bardeen:1972fi,Wald:1974kya}—render its occurrence in natural astrophysical environments highly improbable.

\begin{figure}[htbp]
    \centering
    \centering
    \includegraphics[width=4.5in]{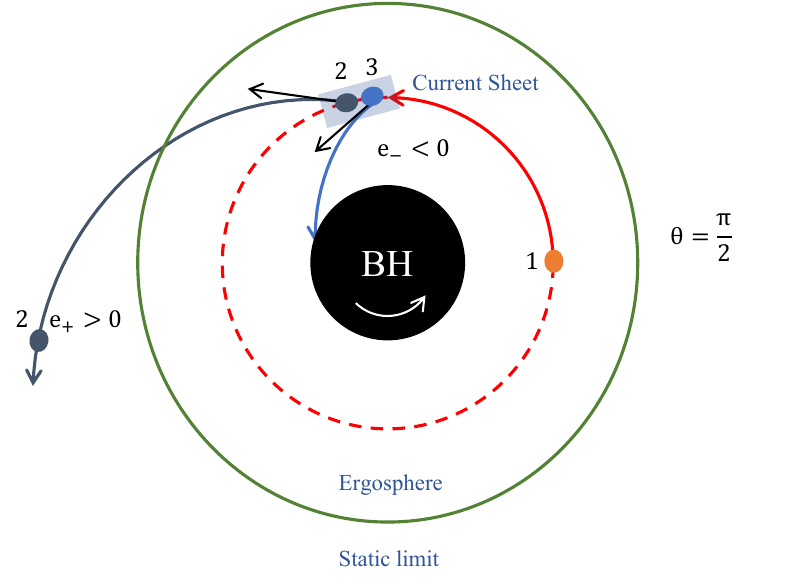}
    \caption{The schematic diagram of the magnetic reconnection-driven Penrose process.}
    \label{mrpp}
\end{figure}

On the other hand, the frame-dragging effect inherent in a rapidly rotating black hole naturally induces a configuration of antiparallel magnetic field lines near the equatorial plane, a phenomenon that aligns with numerical simulations of such rapidly spinning black holes \cite{Parfrey:2018dnc,Komissarov:2005wj,East:2018ayf}. The abrupt transitions in magnetic field lines dynamically generate an equatorial current sheet. When the aspect ratio of this current sheet exceeds a critical threshold, plasma instabilities cause it to break apart—an outcome facilitated by non-ideal magnetohydrodynamic effects such as thermal inertia, pressure torques, or resistivity. Subsequently, the formation of plasma flux ropes drives rapid magnetic reconnection, swiftly converting the available magnetic energy into the kinetic energy of plasma particles. As reconnection proceeds, plasma is ejected from the reconnection layer, accompanied by the release of magnetic tension that drives plasma outflows. Simultaneously, magnetic field lines, subjected to the persistent influence of frame-dragging, are stretched once more, forming new current sheets and sustaining the ongoing reconnection process. During this reconnection process, a fraction of the plasma is accelerated, while another portion is decelerated. As noted in \cite{Comisso:2020ykg}, if this process occurs within the ergosphere, the decelerated plasma may settle into negative-energy orbits and ultimately fall into the black hole, while the accelerated plasma escapes to infinity, thereby extracting energy from the black hole—effectively triggering the Penrose process.

As illustrated in Fig.~\ref{mrpp}, we first assume the existence of a plasmoid, denoted as Plasmoid 1, undergoing Keplerian motion within the ergosphere. Furthermore, we posit that the process of magnetic reconnection, which accelerates the plasma, occurs within a highly localized region that is negligible in scale compared to the black hole's ergosphere. Additionally, we assume that this process transpires instantaneously and that the magnetic field configuration near the black hole does not influence the subsequent trajectory of the plasma.

The four-velocity of Plasmoid 1 is given by  
\begin{equation}
u_K^a = \left( u_K^t, 0, 0, \Omega_K u_K^t \right) \, ,
\label{omegak}
\end{equation}  
where $u_K^t = dt/d\tau$, with $\tau$ being the proper time, and $\Omega_K=d\phi/dt$ is the angular velocity. Considering that Keplerian orbits within the ergosphere can only be prograde, one can obtain from the geodesic equation that  
\bea
u_K^t = \frac{a M^{1/2} + r^{3/2}}{2 a M^{1/2} r^{3/2} + r^2 (-3M + r)} \,,\quad\Omega_K = \frac{M^{1/2}}{r^{3/2} + a M^{1/2}} \, . 
\eea

In order to specifically analyse the localised process of magnetic reconnection, we introduce a locally comoving coordinate system that rotates with the plasma at the Keplerian angular velocity $\Omega_K$ within the equatorial plane. The corresponding tetrads are given by
\bea
\hat \sigma_{(0)}=u^t_K(\partial_t+\Omega_K\partial_\phi),\quad
\hat \sigma_{(1)}=\frac{\partial_r}{\sqrt{g_{rr}}},\quad
\hat \sigma_{(2)}=\frac{\partial_\theta}{\sqrt{g_{\theta\theta}}},\quad
\hat \sigma_{(3)}=\frac{\mu}{g_{tt}}\partial_t+\frac{\nu}{g_{t\phi}}\partial_\phi
\eea
where
\bea
\mu = \frac{g_{t\phi} + g_{\phi\phi} \Omega_K}{D} \, ,\quad
\nu = \frac{-g_{tt} + g_{t\phi} \Omega_K}{D} \, .
\eea
with $D=\left( g_{tt} g_{\phi\phi} - g_{t\phi}^2 \right) \left[ g_{tt} + \Omega_K \left( 2 g_{t\phi} + g_{\phi\phi} \Omega_K \right) \right]$. The direction of the reconnecting magnetic field lines is arbitrary, as it ultimately depends on large-scale magnetic configurations, the rotation of the black hole, and temporal evolution. To specify this direction, we introduce the orientation angle
\[
\xi = \arctan\left(\frac{v^{\prime(1)}_{\text{out}}}{v^{\prime(3)}_{\text{out}}}\right)\,,
\]
where $v^{\prime(1)}_{\text{out}}$ and $v^{\prime(3)}_{\text{out}}$ denote the radial and azimuthal components, respectively, of the plasma outflow velocity emerging from the reconnection layer, as measured in the co-moving frame. Therefore, the velocity of plasma ejected from the reconnection layer can be expressed as
\[
v'_{\pm} = v_{\text{out}} \left( \pm \cos\xi\, \hat{\sigma}^{(3)} \mp \sin\xi\, \hat{\sigma}^{(1)} \right)\,, \label{vlocal}
\]
where $v_{\text{out}}$ refers to the magnitude of the outflow speed observed in the co-moving frame. Here, the subscripts $+$ and $-$ correspond to prograde and retrograde plasma flows, respectively—the former escaping away from, and the latter plunging into, the black hole—and are associated with plasmoids 2 and 3 in Fig.~\ref{mrpp}, respectively. This velocity can be approximated as a function solely of the upstream plasma magnetisation, $\sigma_0 = B_0^2 / \omega$\, where $B_0$ signifies the magnitude of the magnetic field and $\omega$ denotes the plasma enthalpy density. The relationship is given by \cite{Comisso:2020ykg}
\begin{equation}
v_{\text{out}} \simeq \left( \frac{\sigma_0}{1 + \sigma_0} \right)^{1/2} \,.
\label{vout}
\end{equation}
Then, their four-velocities in the BL coordinate system can be expressed as  
\bea  
u^a_\pm = \gamma_{\text{out}} \left( u^t_K \pm \frac{\mu}{g_{tt}} v_{\text{out}} \cos \xi,\, \mp \frac{v_{\text{out}} \sin \xi}{\sqrt{g_{rr}}},\, 0,\, \Omega_K u^t_K \pm \frac{\nu}{g_{t\phi}} v_{\text{out}} \cos \xi \right), \label{uout}  
\eea
with $\gamma_{\text{out}}=\left(1-v^2_{\text{out}}\right)^{-1}$. We assume that magnetic reconnection facilitates a highly efficient conversion of energy, wherein the majority of magnetic energy is transformed into plasma kinetic energy. Consequently, the electromagnetic energy at infinity becomes negligible compared to the hydrodynamic energy of the fluid. Under this assumption, the plasma can be treated as incompressible and adiabatic, allowing an estimate of the energy density of the escaping plasma at infinity as follows \cite{Koide:2008xr}:
\begin{equation}\label{inftye}
e_{\pm} = \omega E_\pm - \frac{p}{u^t_\pm} \,,
\end{equation}
where $p$ denotes the plasma pressure and $E_\pm = -g_{0\mu} u^\mu_\pm$. For a relativistic hot plasma, we adopt a polytropic index of $\Gamma = \frac{4}{3}$\,; hence,
\begin{equation}\label{gamma}
\frac{p}{\omega} = \frac{\Gamma - 1}{\Gamma} = \frac{1}{4} \,.
\end{equation}

Substituting Eqs.~\eqref{uout} and~\eqref{gamma} into Eq.~\eqref{inftye} and performing simplification, we derive the ratio of the hydrodynamic energy to the enthalpy $\epsilon_\pm =e_\pm/\omega$ at infinity as:
\begin{equation}\label{finale}
\begin{split}
\epsilon_\pm = E_K (1 + \sigma_0)^{1/2} \pm (\nu + \mu) \cos\xi \, \sigma_0^{1/2}- \frac{1}{4} \cdot \frac{1}{u^t_K (1 + \sigma_0)^{1/2} \mp \mu (2M / r_0 - 1)^{-1} \cos\xi \, \sigma_0^{1/2}} \,.
\end{split}
\end{equation}
where $E_{K}=u^{t}_{K}\!\left(g_{tt}+g_{t\phi}\,\Omega_{K}\right)$,  which denotes the conserved specific energy of a test particle on a circular Keplerian orbit (i.e., the energy per unit rest mass associated with the Keplerian motion). In addition,  $r_{0}$ denotes the location of the reconnection X-point.

Let us briefly recapitulate this physical process. Initially, the parent plasmoid 1, located within the ergosphere, follows a Keplerian trajectory, moving with a velocity given by Eq.~(\ref{omegak}). At a certain moment, magnetic reconnection occurs instantaneously, causing the parent plasmoid to split into two components, 2 and 3—with 2 escaping to infinity and 3 plunging into the black hole. Their initial four-velocities are determined by Eq.~(\ref{uout}), corresponding to the "+" and "–" branches, respectively.

If the process constitutes a Penrose mechanism induced by magnetic reconnection—that is, rotational energy has been extracted from the black hole—we find that component 2 possesses more energy at infinity than it would if it were at rest at infinity, namely,  
$\Delta \epsilon_+ = \epsilon_+ > 0$\,; and component 3 has negative energy at infinity (i.e.,  
$\epsilon_- < 0$)\,.

\section{Hotspot model and imaging method}\label{sec3}

In this section, we present our hotspot model along with the associated imaging methodology. As our aim is not to investigate the influence of radiative mechanisms on imaging or flaring phenomena, the model does not incorporate the detailed emission spectrum of the hotspot radiation. 
Accordingly, we assume that the hotspot emission is isotropic and frequency-independent, effectively modelling it as a broadband source with a flat spectrum. The plasmoid is treated as a transparent hotspot with an emissivity $J_{\nu_s}$ characterised by a Gaussian distribution of width $s$:
\bea
J_{\nu_s} = e^{-\frac{x^2}{2s^2}} \,,
\eea
where $x$ denotes the distance from the centre of the hotspot. In order to image a moving hotspot, it is essential to track the trajectory of the light source as well as its radiation transfer. According to Eq.~(\ref{uout}), the trajectory of the source is determined via numerical integration of the geodesic equations in their Hamilton–Jacobi form.

For radiation transfer, we employ a numerical backward ray-tracing method combined with a fisheye camera model to generate images. Technical details of the implementation can be found in Appendix B of \cite{Hu:2020usx}. The camera is situated in a Zero Angular Momentum Observer (ZAMO) frame, with corresponding tetrads given by:
\bea
\begin{aligned}
\hat{e}_{(0)} = \frac{g_{\phi\phi} \, \partial_t - g_{t\phi} \, \partial_\phi}{\sqrt{g_{\phi\phi} \left( g_{t\phi}^2 - g_{\phi\phi} g_{tt} \right)}}, \quad \hat{e}_{(1)} = -\frac{\partial_r}{\sqrt{g_{rr}}}\,,\quad\hat{e}_{(2)} = \frac{\partial_\theta}{\sqrt{g_{\theta\theta}}}\,, \quad \hat{e}_{(3)} = -\frac{\partial_\phi}{\sqrt{g_{\phi\phi}}}.
\end{aligned}
\eea
The intensity on the image plane is governed by the radiation transfer equation \cite{Hou:2022eev}:
\bea
\frac{d}{d \lambda}\left(\frac{I_{\nu}}{\nu^{3}}\right) = \frac{J_{\nu}}{\nu^{2}} \,,
\eea
where $\lambda$ denotes the affine parameter along null geodesics, and $I_{\nu}$ and $J_{\nu}$ are the specific intensity and emissivity at frequency $\nu$, respectively. The frequency $\nu = -k^{\mu} u_{\mu}$ is measured by a comoving observer, with $k^{\mu}$ representing the photon's four-momentum. In this analysis, the absorption by the medium is neglected.

Our numerical computations yield a sequence of snapshots, where the photon-arrival time at the image plane is given by the sum of the orbital time and the light-travel time from the hotspot to the observer. Additionally, at each temporal frame, we can determine the flux centroid position traversing the camera plane. According to the definition of the camera employed in \cite{Hu:2020usx}, the flux impinging on the $(i,j)$-th pixel of the screen is given by \cite{Huang:2024wpj}:

\bea
F(i, j) = I_{o} S_{0} \cos \left[2 \arctan \left( \frac{1}{n} \tan \left( \frac{\alpha_{\mathrm{fov}}}{2} \right) \sqrt{\left(i - \frac{n + 1}{2} \right)^2 + \left(j - \frac{n + 1}{2} \right)^2} \right) \right] \,.
\eea
Here, $S_0$ denotes the area of a single pixel, $n$ is the total number of pixels along the horizontal or vertical axis, and $i, j$ range from 1 to $n$. The parameter $\alpha_{\mathrm{fov}}$ stands for the field of view of the camera. Once this data is acquired, the centroid position $\vec{x}_c(t)$ of each image can be evaluated via:
\bea
\vec{x}_c(t) = \frac{\sum_{i,j} \vec{x}(i, j) F(i, j)}{\sum_{i,j} F(i, j)} \,,
\eea
where $\vec{x}(i,j)$ indicates the spatial coordinate of the $(i,j)$-th pixel. The total flux in a given snapshot, denoted by $\sum_{i,j} F(i,j)$, serves as the flux associated with $\vec{x}_c(t)$. Through the use of our model and the aforementioned computational framework, we can ultimately track the temporal evolution of both the brightness centroid’s position and the corresponding flux. Similar methods have also been employed in several other studies; see \cite{Hu:2025hdr, Xie:2025skg, Hu:2025lyp}.

\section{Results and discussion}\label{sec4}

In this section, we present the observational signature of plasmoids undergoing magnetic reconnection via the Penrose process. For simplicity, and without loss of generality, we set the mass of the Kerr black hole to $M = 1$. Considering the substantial cosmic distance separating the black hole from us, we choose the observer's radial coordinate in our numerical simulations to be $r_o = 200 \gg r_+$. The inclination angle of observation is taken as $\theta_o = \pi/10$. Our analysis primarily focuses on two cases, namely $a = 0.94$ and $a = 0.99$, for which the ergoregion lies within the intervals $1.34 < r < 2$ and $1.14 < r < 2$, respectively. Since the splitting occurs within the ergosphere along the equatorial plane, we uniformly set the initial coordinates of the undivided plasmoid to $(r_{s0}, \theta_{s0}, \phi_{s0}) = (1.6, \pi/2, 0)$, and fix its radius at $s = 0.2$. Here we identify $r_{s0}\equiv r_{0}$ as introduced earlier, i.e., the location of the reconnection X-point. The instant when the hotspot first becomes visible on the screen is defined as $t = 0$.

\subsection*{No Magnetic Reconnection: Keplerian Orbital Plasmoid}
\begin{figure}[H]
\centering
\includegraphics[width=0.85\linewidth]{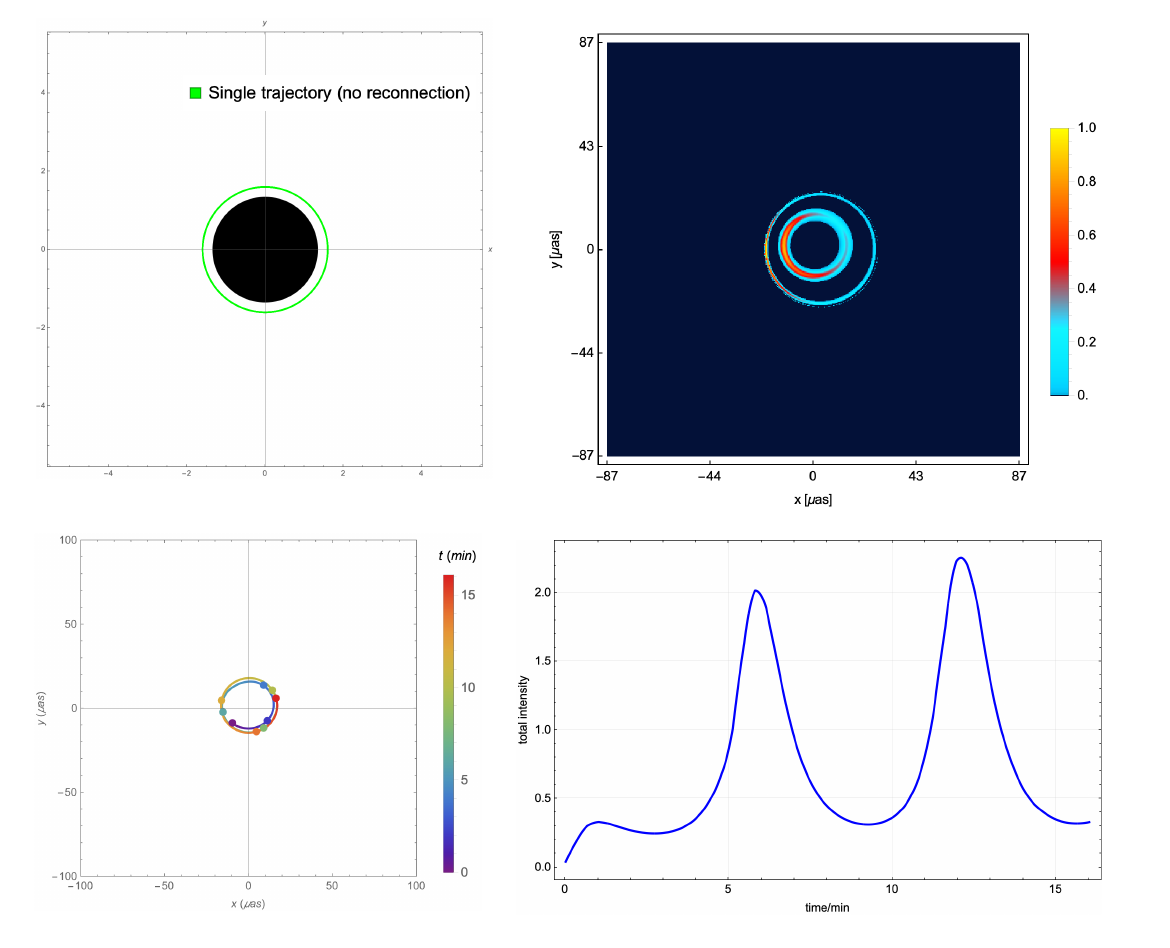}
\caption{Image features in the absence of magnetic reconnection. The top-left panel presents a schematic of the plasmoid's trajectory in a two-dimensional Cartesian coordinate system, where the coordinates are defined as $x = r \sin\theta \cos\phi$, $y = r \sin\theta \sin\phi$. The black circular region in the image represents a Kerr black hole. The top-right panel depicts the intensity distribution of the hotspot as seen by the observer, illustrating the time-averaged radiation intensity across the observational plane; the intensity values are normalized by $I/I_{\mathrm{max}}$\,. The bottom-left panel shows the temporal evolution of the centroid position, with colour encoding observational time (measured in minutes) and a temporal resolution of 2 minutes between successive data points. The bottom-right panel displays the light curve of the hotspot emission, revealing the variation of total flux over observational time.}
\label{fig:img1}
\end{figure}

To facilitate subsequent analysis of the visual characteristics in the magnetic-reconnection-driven Penrose process, we first examine a trivial scenario to serve as a control group: a case in which the initial plasmoid undergoes no magnetic reconnection, remaining intact and stably orbiting the black hole. In this context, the plasmoid maintains a stable circular motion confined to the equatorial plane, exhibiting no energy extraction effects.  This scenario is illustrated in Fig.~\ref{fig:img1}, where the spin parameter of the Kerr spacetime is fixed at $a = 0.94$, the observer's azimuthal angle is specified as $\phi_o = \pi/2$, and the plasmoid is prograde around the black hole.

As depicted in Fig.~\ref{fig:img1}, the time-averaged image of a hotspot undergoing Keplerian motion exhibits a dual-ring structure: the outer ring corresponds to the primary image, while the inner ring represents the secondary image. Within a single orbital period, the light curve features two prominent flares—an observation that aligns with the findings of Ref.~\cite{Huang:2024wpj}, despite the key difference that their study considered an opaque light source, in contrast to the transparent emitter examined in the present work.

\begin{figure}[htbp]
    \centering
    \includegraphics[width=0.85\linewidth]{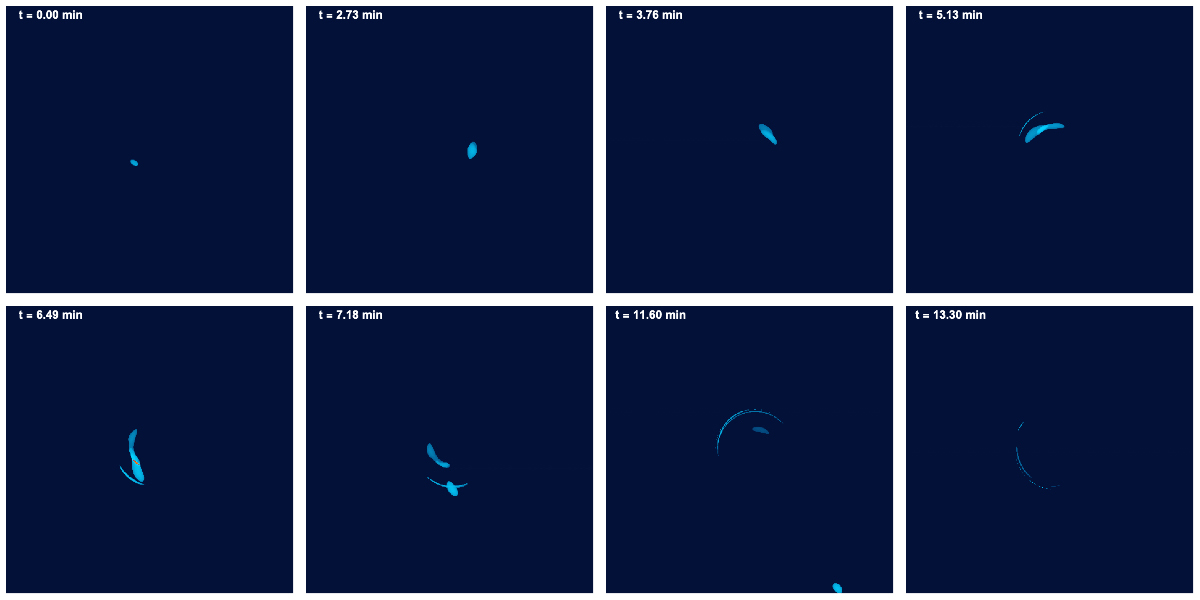}
    \caption{Temporal evolution of the hotspot intensity distribution in the magnetic-reconnection-driven Penrose  process, composed of eight frames showcasing key moments. The timestamp in the upper left corner of each frame indicates the corresponding observational time in minutes. The sequence vividly reveals the dynamic spatial shifts of the hotspot as well as the evolving characteristics of its brightness distribution.}
    \label{fig:split}
\end{figure}

\subsection*{Magnetic Reconnection: Case 1 ($\phi_o=\pi/2$)}

\begin{figure}[H]
    \centering
    \includegraphics[width=0.85\linewidth]{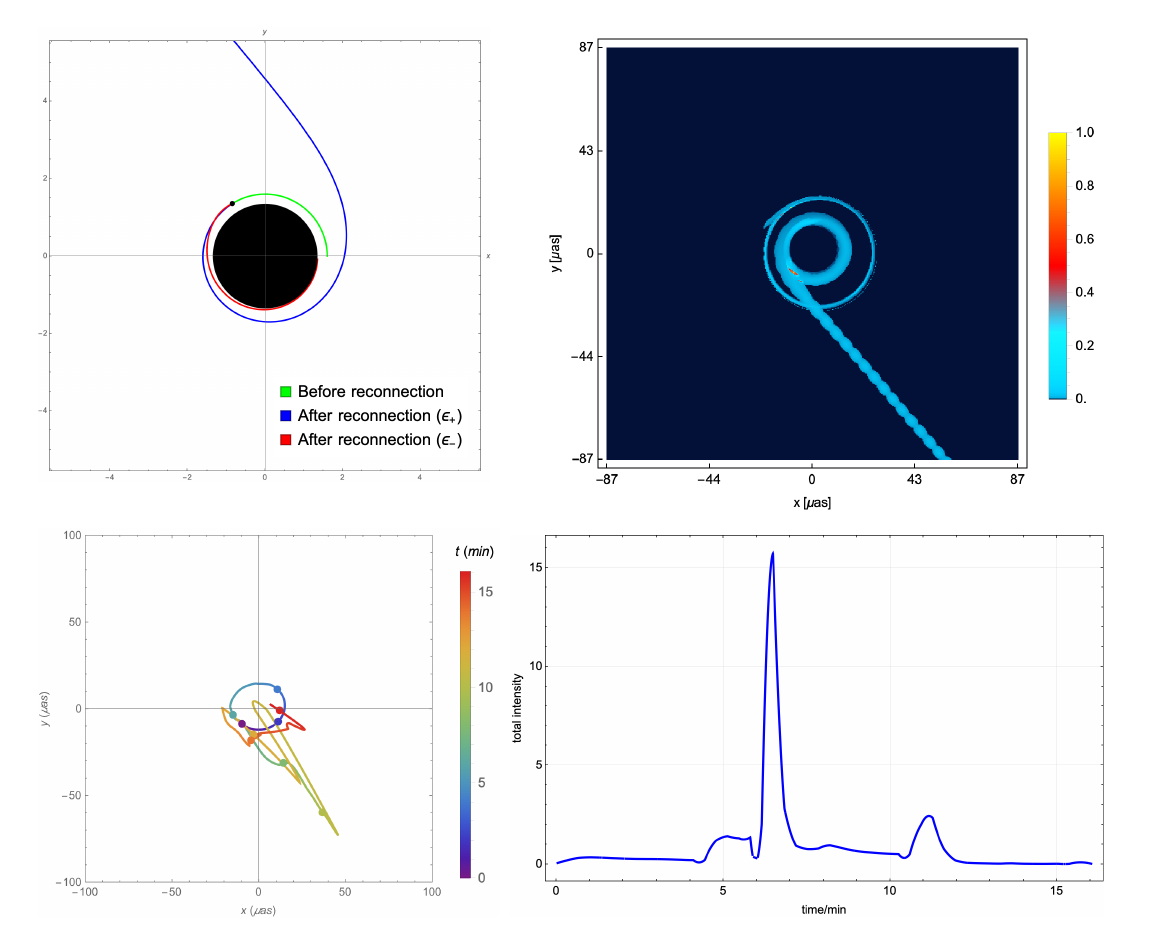}
    \caption{Features of the magnetic-reconnection-driven Penrose process. The upper-left panel presents a schematic diagram of plasmoid trajectories in a two-dimensional Cartesian coordinate system. The black circular region denotes the event horizon of a Kerr black hole. The green solid line represents the trajectory of the plasmoid prior to reconnection, the blue solid line indicates the path of the ejected plasmoid after reconnection, and the red solid line shows the trajectory of the plasmoid falling into the black hole. The black dot marks the location where the reconnection event occurs. The upper-right panel displays the intensity distribution of the hotspot as perceived by a distant observer. The lower-left panel depicts the evolution of the brightness centroid as a function of observational time, with a uniform time interval of two minutes between successive points. The lower-right panel shows the light curve corresponding to the radiation emitted by the hotspot.}
    \label{fig:a94mrpp}
\end{figure}

In this case, we choose a magnetization parameter of $\sigma_0 = 25$ and set the orientation angle to $\xi = \pi / 20$, with the spin parameter of the Kerr spacetime fixed at $a = 0.94$, and the observer's azimuthal angle specified as $\phi_o = \pi/2$. According to Eq.~\eqref{finale}, we obtain $\epsilon_+ \approx 9.89 > 0$ and $\epsilon_- \approx -0.50 < 0$, thus satisfying both criteria for energy extraction. It follows that under these conditions, rotational energy has indeed been extracted from the black hole. The results for this example are presented in Figs.~\ref{fig:split} and~\ref{fig:a94mrpp}.

Fig.~\ref{fig:split} illustrates the temporal evolution of hotspot intensity distributions during the magnetic-reconnection-driven Penrose  process. At time $t = 0$, a plasmoid that has yet to undergo magnetic reconnection is observed on the screen.  
At $t = 2.73$, the moment of reconnection—which is essentially instantaneous—has just occurred. Upon close inspection, one can discern that the blue spot visible in the image consists of several superimposed blue points, though this is not immediately apparent. By $t = 3.76$, two distinct plasmoids can be clearly identified on the screen.  
At $t = 5.13$, the plasmoid plunging into the black hole (on the right) reaches its peak brightness, corresponding to the 5-minute flare observed in the light curve of Fig.~\ref{fig:a94mrpp}. At $t = 6.49$, the escaping plasmoid attains its maximum luminosity; the red region corresponds to the most prominent flare at 6 minutes on the same light curve. At $t = 7.18$, the two plasmoids are fully separated: one can clearly see a plasmoid beginning its journey toward spatial infinity, while the other remains bound near the black hole. At $t = 11.60$, the escaping plasmoid is about to exit the visible screen, and the infalling plasmoid has nearly vanished into the black hole. By $t = 13.3$, the primary images of both plasmoids have completely disappeared, leaving only residual secondary and higher-order images faintly lingering on the screen.

\begin{figure}[H]
    \centering
    \includegraphics[width=0.85\linewidth]{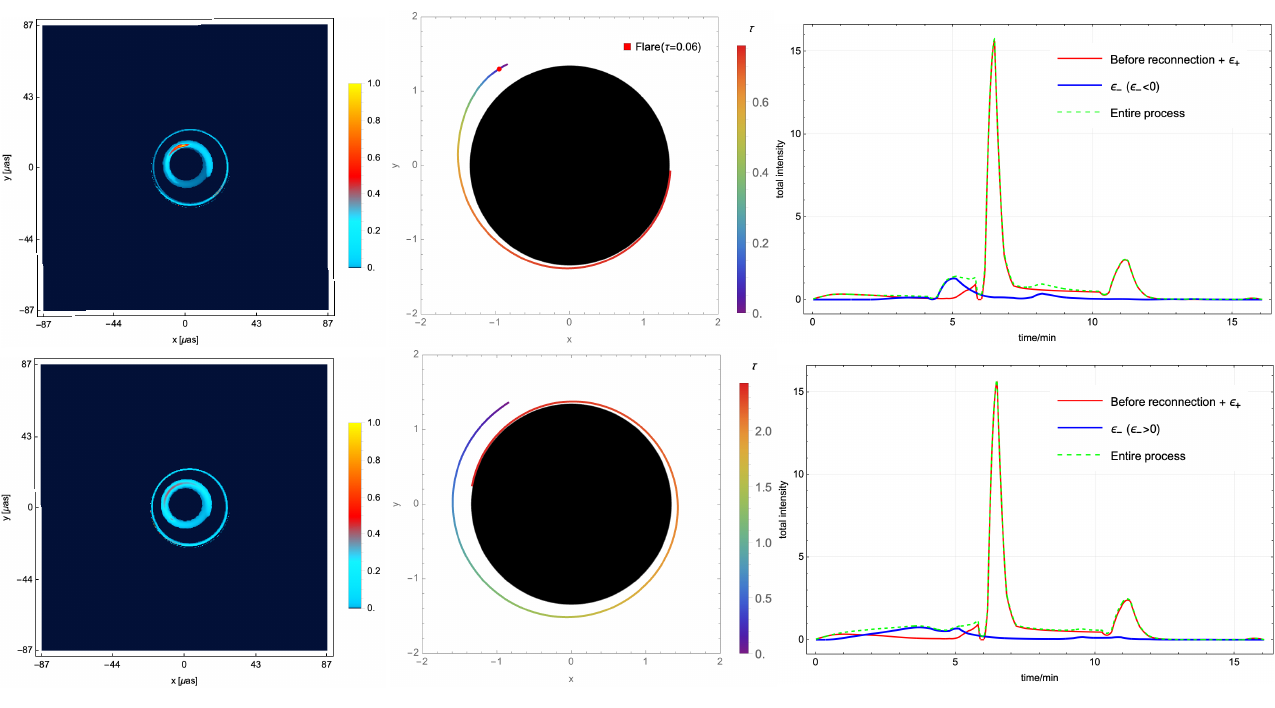}
    \caption{Comparison plots of the intensity distribution for positive- and negative-energy plasmoids plunging into the black hole are shown. The upper row corresponds to negative-energy plasmoids with $\epsilon_{-} \approx -0.50$, while the lower row depicts positive-energy plasmoids with $\epsilon_{+} \approx 0.50$. The first column presents the normalized intensity distributions generated solely by the plasmoids as they fall into the black hole. The second column displays schematic trajectories of the infalling plasmoids. The color gradient along each trajectory illustrates the evolution of the plasmoid's position as a function of its proper time $\tau$. Specifically, we mark the position corresponding to approximately $\tau = 0.06$ in the plot at the centre of the first row, which precisely identifies the origin of the flare within the underlying light curves. The third column offers the light curves arising from hotspot emission, providing a comparative view of the total flux versus time for three scenarios: pre-reconnection and escaping plasmoids (solid red curve), infalling plasmoids (solid blue curve), and the overall process (dashed green curve).}
    \label{fig:a94pm}
\end{figure}  

In the lower-right panel of Fig.~\ref{fig:a94mrpp}, the corresponding light curve is shown, illustrating the variation of the total radiative flux from the hotspot as perceived over observational time. A marked surge in flux appears around the seventh minute, signifying the emergence of a flare. Furthermore, two relatively weaker flares emerge near the fifth and eleventh minutes, respectively.

Next, we aim to thoroughly analyze the underlying causes of the three observed flares. In particular, since we seek to determine from the light curve whether a Penrose process has taken place or not, it is crucial that the light curve capture distinctive signatures indicative of such a phenomenon. One of the most essential prerequisites for the occurrence of a Penrose process is the infall of negative-energy plasmoids into the black hole. If such characteristic signals can be extracted from the light curve, it would hold significant physical implications.

Therefore, in the upper-right panel of Fig.~\ref{fig:a94pm}, we illustrate the individual light curves of the positive-energy plasmoid that escapes to infinity and the negative-energy plasmoid that plunges into the black hole, both following magnetic reconnection. For convenience, we juxtapose the pre-reconnection light curve with that of the positive-energy component, representing it with a solid red curve. The light curve corresponding to the negative-energy plasmoid is represented by a solid blue line, while the total light curve is depicted with a dashed green line. To underscore the distinctive features of the light curve produced by the negative-energy plasmoid with $\epsilon_{-} \approx -0.50$, we artificially replace it with a positive-energy plasmoid of $\epsilon_{-} \approx 0.50$, keeping the orientation angle at $\xi = \pi / 20$ and preserving the trajectory and energy of the escaping plasmoid, and subsequently compute the resulting light curve shown in the lower-right panel of Fig.~\ref{fig:a94pm}. The interpretations of the red solid, blue solid, and green dashed curves remain consistent with the previously established convention. Furthermore, in the central panel of Fig.~\ref{fig:a94pm}, we present the trajectories and corresponding proper times required for both the negative- and positive-energy plasmoids as they spiral into the black hole.

As shown in the upper-right panel of Fig.~\ref{fig:a94pm}, the most prominent flare, appearing around the 7th minute, as well as the comparatively weaker flare, around the 11th minute in the light curve, are both induced by the escaping plasmoid. This phenomenon arises as a result of the escaping plasmoid acquiring substantial kinetic energy and undergoing significant Doppler blueshift, along with its secondary image also being subject to Doppler blueshift effects. The relatively weaker flare occurring around the 5th minute is triggered by the negative-energy plasmoid plunging into the black hole, arising from the Doppler blueshift of its primary image. A more thorough discussion of the physical mechanisms underlying flare formation is provided in Appendix \ref{App:detail}.  As shown in the lower-right panel of Fig.~\ref{fig:a94pm}, the equivalent positive-energy plasmoid falling into the black hole does not generate such a flare. Moreover, given that the negative-energy plasmoid rapidly falls into the black hole, it has sufficient time to produce only a single flare. Taken together, these observations suggest that this flare feature is distinctive under such circumstances and may serve as a potential observational signature of a Penrose process in action.

\subsection*{Magnetic Reconnection: Case 2 ($\phi_o=0$)}

\begin{figure}[H]
    \centering
    \includegraphics[width=0.85\linewidth]{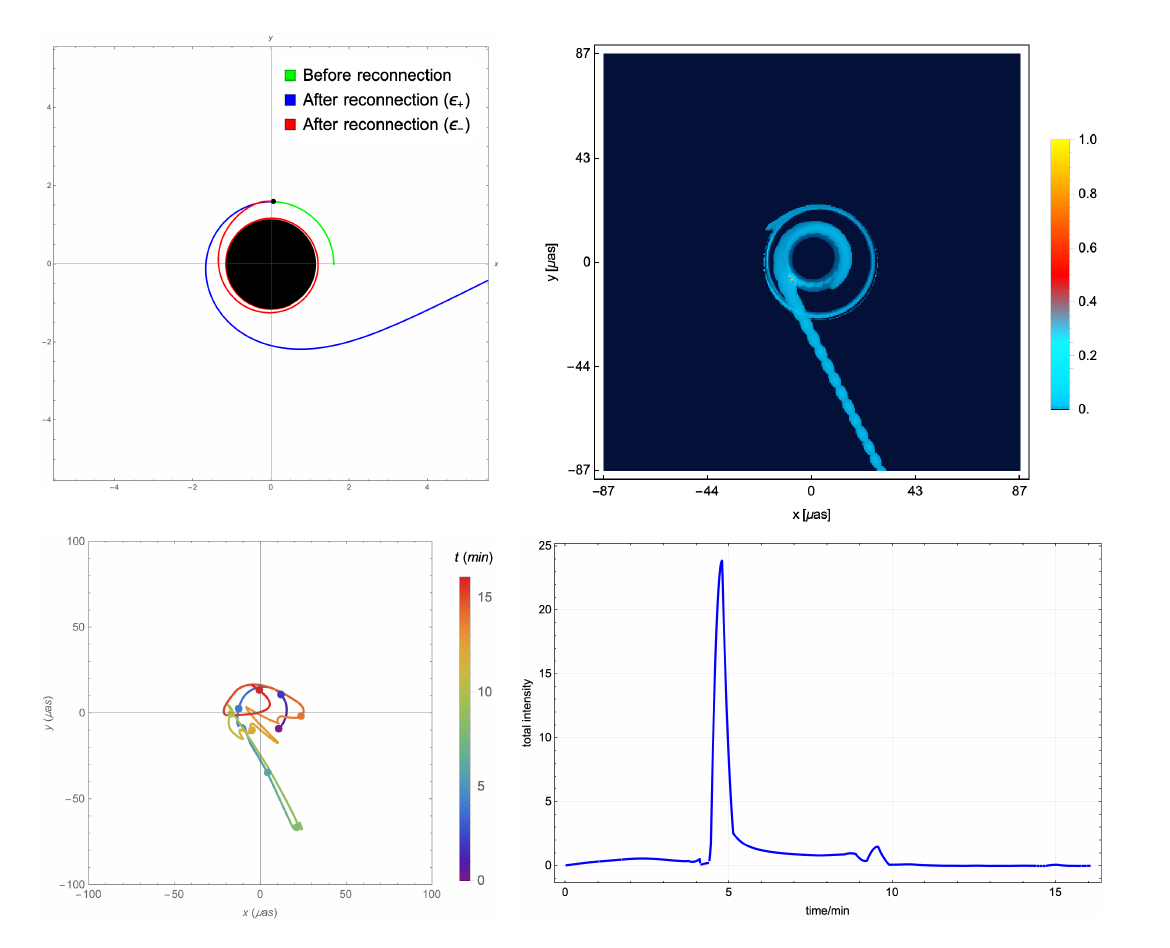}
    \caption{For spin parameter $a = 0.99$, the imaging signatures of the magnetic reconnection–induced Penrose process are presented as follows: the top-left panel illustrates the motion of the plasmoid within a two-dimensional Cartesian coordinate system; the top-right panel reveals the spatial distribution of intensity from the hotspot as perceived by a distant observer; the bottom-left panel delineates the temporal evolution of the brightness centroid's location throughout the observation;  
while the bottom-right panel portrays the light curve of the hotspot’s radiation, capturing its flux variations over time.}
    \label{fig:a99}
\end{figure}

\begin{figure}[h]
    \centering
    \includegraphics[width=1\linewidth]{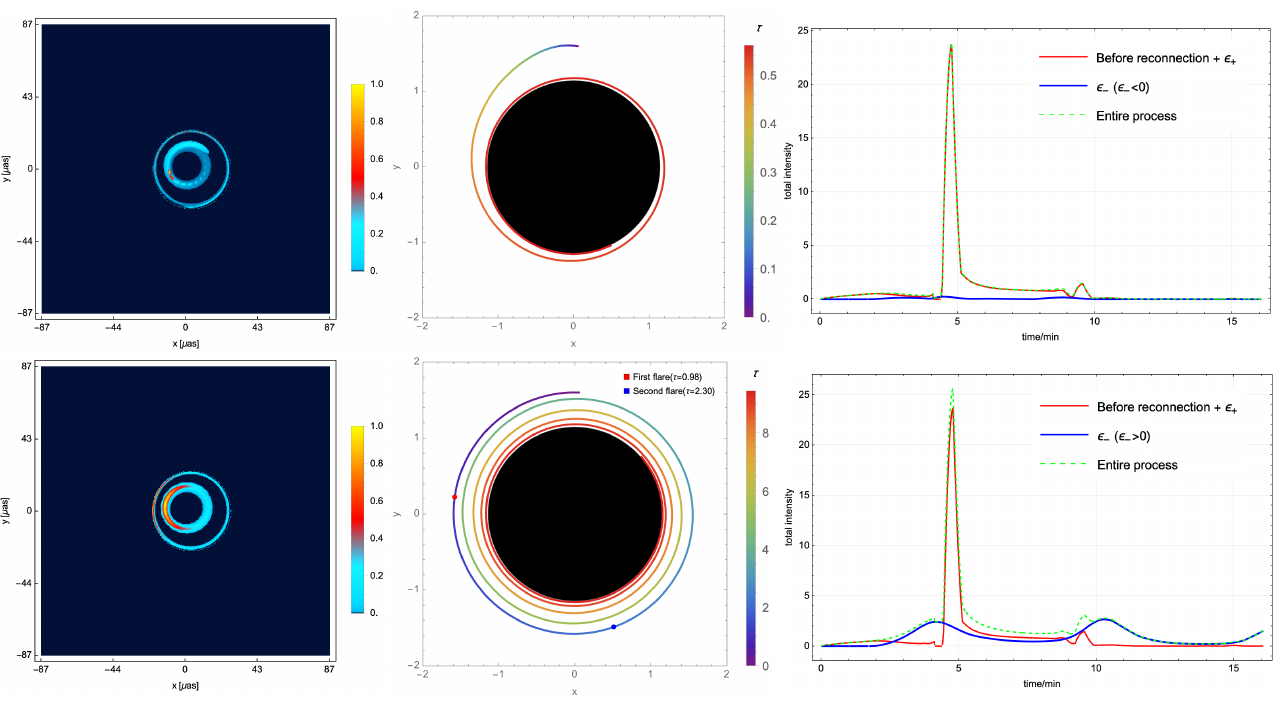}
    \caption{A comparative visualization showcasing the intensity distributions associated with the infall of negative- and positive-energy plasmoids into a black hole is presented. The first row corresponds to negative-energy plasmoids, characterized by $\epsilon_- \approx -0.66$, while the second row pertains to positive-energy plasmoids, with $\epsilon_- \approx 0.66$. The first column illustrates the normalized intensity distributions generated exclusively by the infalling plasmoids. The second column provides schematic depictions of the trajectories followed by the plasmoids as they descend into the black hole. The colour along each trajectory indicates the evolution of the plasmoid's position with respect to its proper time $\tau$. Notably, the positions corresponding to approximately $\tau = 0.98$ and $\tau = 2.30$ are marked within the right-hand panels; these two points coincide with the origins of the twin flares observed in the lower blue light curve. The third column presents the light curves of radiation emitted by the hotspots.}
    \label{fig:a99pm}
\end{figure}   

Previously, we considered the scenario in which a negative-energy plasmoid plunging into a black hole generates a faint flare, while a positive-energy plasmoid with the same orientation angle fails to produce any noticeable flare. We must extend our analysis to examine the possibility that a positive-energy plasmoid falling into the black hole might, in certain cases, also produce a flare. Under such circumstances, it becomes crucial to investigate whether the light curve generated by a negative-energy plasmoid with the same orientation angle would exhibit distinct characteristics compared to that of its positive-energy counterpart. Should such differences exist, they could likewise serve as compelling observational signatures indicative of a possible Penrose process. Let us now proceed to explore this scenario.

In this scenario, we select a spin parameter of $a = 0.99$, while keeping the magnetization parameter fixed at $\sigma_0 = 25$ and the orientation angle at $\xi = \pi / 20$. To intentionally obtain a phenomenon distinct from that in Case 1—namely, a flare triggered by the infall of a positive-energy plasmoid into the black hole—we set the observer’s azimuthal angle to $\phi_o = 0$. Substituting into equation~\eqref{finale}, we obtain the energy of the escaping plasmoid as $\epsilon_+ \approx 8.21 > 0$, and the energy of the infalling plasmoid as $\epsilon_- \approx -0.66 < 0$. This configuration still satisfies the condition for energy extraction. The results are illustrated in Fig.~\ref{fig:a99}.

Similarly, to gain a clearer understanding of the specific origin of the flare depicted in the lower right panel of Fig.~\ref{fig:a99}, and, more importantly, to determine whether it is possible in this scenario to extract from the light curve key signatures that might distinctly indicate the occurrence of a Penrose process, we extend the methodology adopted in the preceding analysis. To this end, a comparative experiment is devised, wherein plasmoids bearing both positive and negative energies are permitted to plunge into the black hole. The energy of the positive-energy plasmoid is manually set to $\epsilon_- \approx 0.66 > 0$, corresponding to the negative-energy case of $\epsilon_- \approx -0.66$, while all other conditions remain consistent with those of the previous setup. The results are presented in Fig.~\ref{fig:a99pm}.

By comparing the right panels in the first and second rows of Fig.~\ref{fig:a99pm}, we observe that the most prominent flare around 5 minutes, along with a smaller flare occurring shortly before 10 minutes, is produced by plasmoids that escape from the black hole. The negative-energy plasmoid that plunges into the black hole generates no flare, whereas the positive-energy plasmoid plunging into the black hole produces two distinct flares—one shortly before 5 minutes and another shortly after 10 minutes. 

This behavior stands in marked contrast to Case 1, where the negative-energy plasmoid gave rise to a single flare, while the positive-energy plasmoid produced none. In Case 2, although the negative-energy plasmoid fails to produce a flare, the positive-energy plasmoid—unlike its negative counterpart that plunges swiftly— orbits the black hole several times before its eventual infall, thereby giving rise to two flares. This constitutes a distinctive feature, which may serve as a potential observational signature of the Penrose process.

\section{Summary}\label{sec5}

In this study, we investigated the Penrose process driven by magnetic reconnection, focusing on plasmoids in Keplerian motion within the ergosphere. Upon energization through reconnection, a plasmoid splits into a positive-energy fragment that escapes to infinity and a negative-energy fragment that plunges into the black hole. By modeling the plasmoid dynamics before and after splitting within a hotspot framework, we obtained the resulting imaging signatures and identified distinctive flare features in the light curves. In particular, we performed a comparative analysis of the flares produced by the infall of positive- and negative-energy plasmoids to identify the flare signatures associated with the negative-energy plasmoid, which serves as the most crucial indicator of the occurrence of the Penrose process.

For a fixed polar viewing angle $\theta_o = \pi/10$, observers located at different azimuthal positions perceive distinct hotspot images and light curves. In this work, we considered two representative cases: one with a black hole spin of $a = 0.94$ and an observer azimuth of $\phi_{o} = \pi / 2$, and the other with a spin of $a = 0.99$ and an observer azimuth of $\phi_{o} = 0$. Our detailed calculations for these two configurations revealed pronounced asymmetric features. Specifically, in the first case, the infall of the negative-energy fragment produced a single flare, whereas the positive-energy fragment under identical conditions yielded none. In the second case, corresponding to a different azimuthal perspective, the negative-energy fragment generated no flare, while the positive-energy one produced two. These contrasting manifestations provide clear observational imprints that can be directly attributed to the presence of negative-energy matter—signatures of the Penrose process in action.

We note that our findings, based on the two specific configurations presented here, do not preclude the possibility that other mechanisms could produce similar flare signatures, nor do they rule out the existence of additional features in a broader parameter study. Thus, more comprehensive and systematic investigations are warranted. Furthermore, while this work considers a single plasmoid, extending this approach to the collisional Penrose process presents a promising direction for future research.

\section*{Acknowledgments}
We are grateful to Jiewei Huang and Zhenyu Zhang for insightful discussions. The work is partly supported by NSFC Grant No. 12575048, 12275004 and 12205013. M. Guo and X. Wang are also supported by Open Fund of Key Laboratory of Multiscale Spin Physics (Ministry of Education), Beijing Normal University. Z.Y. Fan was supported in part by the National Natural Science Foundations of China with Grant No. 11805041 and No. 11873025 and also supported in part by Guangzhou Science and Technology Project 2023A03J0016.

\appendix
\appendixtrue  % 开启附录样式

\section{A detailed analysis of the flare formation mechanism}
\label{App:detail}

In this appendix, we undertake a more detailed analysis of the mechanism by which plasmoids—such as those illustrated in Figs. \ref{fig:a94pm} and \ref{fig:a99pm}—plunge into a black hole and give rise to flares, considering both Scenarios 1 and 2 within the process of magnetic reconnection.

\begin{figure}[h]
    \centering
    \includegraphics[width=1\linewidth]{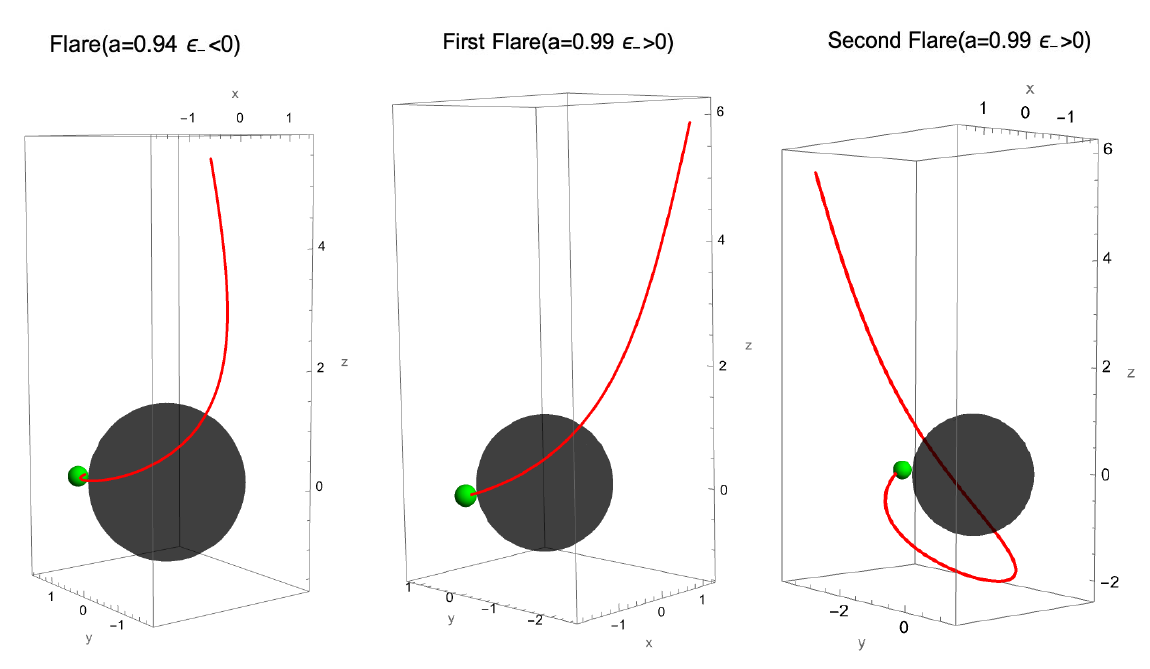}
    \caption{The photon trajectories corresponding to flares generated by plasmoids plunging into the black hole, with the left panel representing Case 1, and the central and right panels corresponding to Case 2.}
    \label{fig:flare}
\end{figure}  

\begin{figure}[h]
    \centering
    \includegraphics[width=1\linewidth]{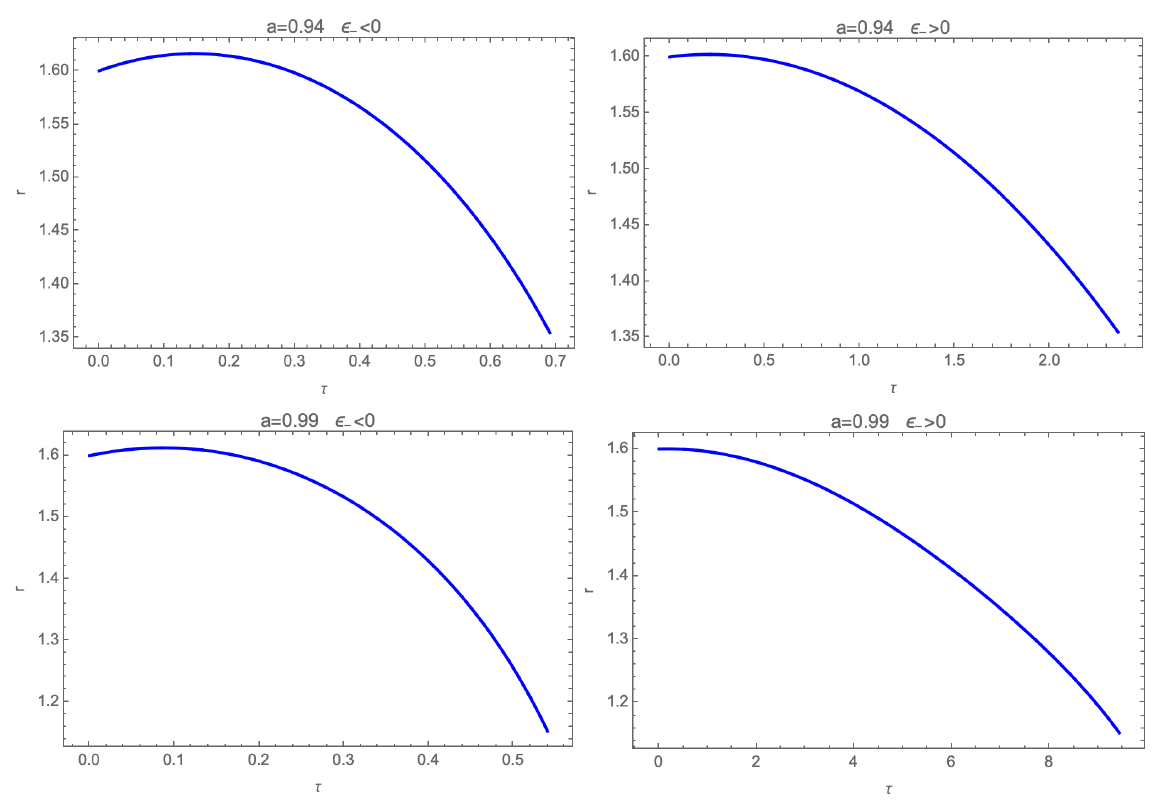}
    \caption{The plots of the radial coordinate $r$ of positive- and negative-energy plasmoids falling into the black hole as functions of the proper time $\tau$, with the first row corresponding to Case 1 and the second row to Case 2.}
    \label{fig:randtau}
\end{figure}  

In Fig. \ref{fig:flare}, we present the photon trajectories corresponding to the flares generated by plasmoids plunging into the black hole in both scenarios. Since the plasmoid moves within the equatorial plane, the initial polar angle of the photons is $\theta_s = \pi/2$. If the photon reaches the observer without crossing the equatorial plane, it forms a primary image; if it crosses the equatorial plane once before reaching the observer, it forms a secondary image \cite{Huang:2024wpj}. From Fig. \ref{fig:flare}, we observe that, in the first two plots, the photons travel directly from the source on the equatorial plane to the observer without crossing the plane. In contrast, in the third plot, the photons, emitted from the equatorial source, traverse the equatorial plane once before reaching the observer. Consequently, it is evident that the flare arising from the negative-energy plasmoid in Fig. \ref{fig:a94pm} is generated by the primary image, while the two flares produced by the positive-energy plasmoid in Fig. \ref{fig:a99pm} originate from the primary and secondary images, respectively.

In Fig. \ref{fig:randtau}, we also present the functional dependence of the radial coordinate on proper time for plasmoids with both positive and negative energies falling into the black hole in Case 1 and Case 2. We observe that in the top row and in the left panel of the second row—that is, for positive and negative energy plasmoids in Case 1, as well as for the negative energy plasmoid in Case 2—turning points appear shortly after departure. By contrast, in the right panel of the second row, corresponding to the positive energy plasmoid in Case 2, no such turning point arises. However, during an initial interval, the radial coordinate remains nearly constant, akin to being in the vicinity of a turning point. A common feature in these scenarios is that when the radial variation is very small, the plasmoid undergoes a pronounced deflection. Consequently, there might arise a moment when its velocity vector is directed, or nearly directed, toward the observer, producing a Doppler blueshift that manifests as a flare. In our previous work \cite{Huang:2024wpj}, we referred to this phenomenon as the Turning Doppler Blueshift Flare (TDBF). Furthermore, from Fig. \ref{fig:a99pm}, we note that in Case 2, the positive-energy plasmoid orbits the black hole several times before plunging, which consequently gives rise to flares induced by secondary images. In our previous work \cite{Huang:2024wpj}, we referred to this phenomenon as the Turning Doppler Blueshift Flare (TDBF). In the present study, we additionally designate the flare triggered by the Doppler blueshift of the secondary image as the Looping Doppler Blueshift Flare (LDBF).

By examining Figs. \ref{fig:a94pm}, \ref{fig:a99pm}, \ref{fig:flare}, and \ref{fig:randtau}, we can infer that in Case 1, a negative-energy plasmoid plunging into the black hole gives rise precisely to a TDBF, whereas a positive-energy plasmoid fails to produce any corresponding flare. In Case 2, however, the positive-energy plasmoid generates both an initial TDBF and subsequently an LDBF, while the negative-energy plasmoid produces no flare at all. From this, we may further draw a conclusion—perhaps not universal in scope—that a negative-energy plasmoid can give rise to a single TDBF or none at all, whereas a positive-energy plasmoid will either produce no flare or else exhibit a TDBF followed by an LDBF. This distinction arises from the fact that positive-energy plasmoids orbit the black hole several times before plunging, while negative-energy plasmoids plunge much more swiftly.

\bibliographystyle{utphys}
\bibliography{reference}

\providecommand{\href}[2]{#2}\begingroup\raggedright\begin{thebibliography}{10}

\bibitem{Penrose:1969pc}
R.~Penrose, ``{Gravitational collapse: The role of general relativity},'' \href{http://dx.doi.org/10.1023/A:1016578408204}{{\em Riv. Nuovo Cim.} {\bfseries 1} (1969) 252--276}.

\bibitem{Penrose:1971uk}
R.~Penrose and R.~M. Floyd, ``{Extraction of rotational energy from a black hole},'' \href{http://dx.doi.org/10.1038/physci229177a0}{{\em Nature} {\bfseries 229} (1971) 177--179}.

\bibitem{Denardo:1974qis}
G.~Denardo, L.~Hively, and R.~Ruffini, ``{On the generalized ergosphere of the Kerr-Newman geometry},'' \href{http://dx.doi.org/10.1016/0370-2693(74)90557-7}{{\em Phys. Lett. B} {\bfseries 50} (1974) 270--272}.

\bibitem{Schnittman:2018ccg}
J.~D. Schnittman, ``{The Collisional Penrose Process},'' \href{http://dx.doi.org/10.1007/s10714-018-2373-5}{{\em Gen. Rel. Grav.} {\bfseries 50} no.~6, (2018) 77}, \href{http://arxiv.org/abs/1910.02800}{{\ttfamily arXiv:1910.02800 [astro-ph.HE]}}.

\bibitem{Banados:2009pr}
M.~Banados, J.~Silk, and S.~M. West, ``{Kerr Black Holes as Particle Accelerators to Arbitrarily High Energy},'' \href{http://dx.doi.org/10.1103/PhysRevLett.103.111102}{{\em Phys. Rev. Lett.} {\bfseries 103} (2009) 111102}, \href{http://arxiv.org/abs/0909.0169}{{\ttfamily arXiv:0909.0169 [hep-ph]}}.

\bibitem{Zhang:2018gpn}
M.~Zhang, J.~Jiang, Y.~Liu, and W.-B. Liu, ``{Collisional Penrose process of charged spinning particles},'' \href{http://dx.doi.org/10.1103/PhysRevD.98.044006}{{\em Phys. Rev. D} {\bfseries 98} no.~4, (2018) 044006}.

\bibitem{Guo:2016vbt}
M.~Guo and S.~Gao, ``{Kerr black holes as accelerators of spinning test particles},'' \href{http://dx.doi.org/10.1103/PhysRevD.93.084025}{{\em Phys. Rev. D} {\bfseries 93} no.~8, (2016) 084025}, \href{http://arxiv.org/abs/1602.08679}{{\ttfamily arXiv:1602.08679 [gr-qc]}}.

\bibitem{Zhang:2016btg}
Y.-P. Zhang, B.-M. Gu, S.-W. Wei, J.~Yang, and Y.-X. Liu, ``{Charged spinning black holes as accelerators of spinning particles},'' \href{http://dx.doi.org/10.1103/PhysRevD.94.124017}{{\em Phys. Rev. D} {\bfseries 94} no.~12, (2016) 124017}, \href{http://arxiv.org/abs/1608.08705}{{\ttfamily arXiv:1608.08705 [gr-qc]}}.

\bibitem{Press:1972zz}
W.~H. Press and S.~A. Teukolsky, ``{Floating Orbits, Superradiant Scattering and the Black-hole Bomb},'' \href{http://dx.doi.org/10.1038/238211a0}{{\em Nature} {\bfseries 238} (1972) 211--212}.

\bibitem{Pani:2012vp}
P.~Pani, V.~Cardoso, L.~Gualtieri, E.~Berti, and A.~Ishibashi, ``{Black hole bombs and photon mass bounds},'' \href{http://dx.doi.org/10.1103/PhysRevLett.109.131102}{{\em Phys. Rev. Lett.} {\bfseries 109} (2012) 131102}, \href{http://arxiv.org/abs/1209.0465}{{\ttfamily arXiv:1209.0465 [gr-qc]}}.

\bibitem{Blandford:1977ds}
R.~D. Blandford and R.~L. Znajek, ``{Electromagnetic extractions of energy from Kerr black holes},'' \href{http://dx.doi.org/10.1093/mnras/179.3.433}{{\em Mon. Not. Roy. Astron. Soc.} {\bfseries 179} (1977) 433--456}.

\bibitem{Bardeen:1972fi}
J.~M. Bardeen, W.~H. Press, and S.~A. Teukolsky, ``{Rotating black holes: Locally nonrotating frames, energy extraction, and scalar synchrotron radiation},'' \href{http://dx.doi.org/10.1086/151796}{{\em Astrophys. J.} {\bfseries 178} (1972) 347}.

\bibitem{Wald:1974kya}
R.~M. Wald, ``{Energy Limits on the Penrose Process},'' \href{http://dx.doi.org/10.1086/152959}{{\em Astrophys. J.} {\bfseries 191} (1974) 231}.

\bibitem{Koide:2008xr}
S.~Koide and K.~Arai, ``{Energy Extraction from a Rotating Black Hole by Magnetic Reconnection in Ergosphere},'' \href{http://dx.doi.org/10.1086/589497}{{\em Astrophys. J.} {\bfseries 682} (2008) 1124}, \href{http://arxiv.org/abs/0805.0044}{{\ttfamily arXiv:0805.0044 [astro-ph]}}.

\bibitem{Asenjo:2017gsv}
F.~A. Asenjo and L.~Comisso, ``{Relativistic Magnetic Reconnection in Kerr Spacetime},'' \href{http://dx.doi.org/10.1103/PhysRevLett.118.055101}{{\em Phys. Rev. Lett.} {\bfseries 118} no.~5, (2017) 055101}, \href{http://arxiv.org/abs/1701.03669}{{\ttfamily arXiv:1701.03669 [astro-ph.HE]}}.

\bibitem{Zeng:2025vjt}
X.-X. Zeng and K.~Wang, ``{Energy extraction via magnetic reconnection in Kerr-Sen-AdS4 black hole: Circular plasma and plunging plasma},'' \href{http://dx.doi.org/10.1103/mnpg-xbc4}{{\em Phys. Rev. D} {\bfseries 112} no.~6, (2025) 064080}, \href{http://arxiv.org/abs/2507.10520}{{\ttfamily arXiv:2507.10520 [gr-qc]}}.

\bibitem{Wang:2025pqh}
K.~Wang and X.-X. Zeng, ``{Energy extraction from the accelerating Kerr black hole via magnetic reconnection in the plunging region and circular orbit region},'' \href{http://arxiv.org/abs/2508.11934}{{\ttfamily arXiv:2508.11934 [gr-qc]}}.

\bibitem{Comisso:2020ykg}
L.~Comisso and F.~A. Asenjo, ``{Magnetic Reconnection as a Mechanism for Energy Extraction from Rotating Black Holes},'' \href{http://dx.doi.org/10.1103/PhysRevD.103.023014}{{\em Phys. Rev. D} {\bfseries 103} no.~2, (2021) 023014}, \href{http://arxiv.org/abs/2012.00879}{{\ttfamily arXiv:2012.00879 [astro-ph.HE]}}.

\bibitem{Wei:2022jbi}
S.-W. Wei, H.-M. Wang, Y.-P. Zhang, and Y.-X. Liu, ``{Effects of tidal charge on magnetic reconnection and energy extraction from spinning braneworld black hole},'' \href{http://dx.doi.org/10.1088/1475-7516/2022/04/050}{{\em JCAP} {\bfseries 04} no.~04, (2022) 050}, \href{http://arxiv.org/abs/2201.12729}{{\ttfamily arXiv:2201.12729 [gr-qc]}}.

\bibitem{Li:2023htz}
Z.~Li and F.~Yuan, ``{Energy extraction via Comisso-Asenjo mechanism from rotating hairy black hole},'' \href{http://dx.doi.org/10.1103/PhysRevD.108.024039}{{\em Phys. Rev. D} {\bfseries 108} no.~2, (2023) 024039}, \href{http://arxiv.org/abs/2304.12553}{{\ttfamily arXiv:2304.12553 [gr-qc]}}.

\bibitem{Chen:2024ggq}
B.~Chen, Y.~Hou, J.~Li, and Y.~Shen, ``{Energy extraction from a Kerr black hole via magnetic reconnection within the plunging region},'' \href{http://dx.doi.org/10.1103/PhysRevD.110.063003}{{\em Phys. Rev. D} {\bfseries 110} no.~6, (2024) 063003}, \href{http://arxiv.org/abs/2405.11488}{{\ttfamily arXiv:2405.11488 [gr-qc]}}.

\bibitem{Fan:2024fcy}
Z.-Y. Fan, Y.~Li, F.~Zhou, and M.~Guo, ``{Fast magnetic reconnection in Kerr spacetime},'' \href{http://dx.doi.org/10.1103/PhysRevD.110.104044}{{\em Phys. Rev. D} {\bfseries 110} no.~10, (2024) 104044}, \href{http://arxiv.org/abs/2409.05434}{{\ttfamily arXiv:2409.05434 [astro-ph.HE]}}.

\bibitem{Khodadi:2022dff}
M.~Khodadi, ``{Magnetic reconnection and energy extraction from a spinning black hole with broken Lorentz symmetry},'' \href{http://dx.doi.org/10.1103/PhysRevD.105.023025}{{\em Phys. Rev. D} {\bfseries 105} no.~2, (2022) 023025}, \href{http://arxiv.org/abs/2201.02765}{{\ttfamily arXiv:2201.02765 [gr-qc]}}.

\bibitem{Shen:2024sdr}
Y.~Shen, H.-Y. YuChih, and B.~Chen, ``{Energy extraction from a rotating black hole via magnetic reconnection: The plunging bulk plasma and orientation angle},'' \href{http://dx.doi.org/10.1103/PhysRevD.110.123010}{{\em Phys. Rev. D} {\bfseries 110} no.~12, (2024) 123010}, \href{http://arxiv.org/abs/2409.07345}{{\ttfamily arXiv:2409.07345 [gr-qc]}}.

\bibitem{Jiang:2024vgn}
H.-X. Jiang, I.~K. Dihingia, C.~Liu, Y.~Mizuno, and T.~Zhu, ``{Enhanced Blandford Znajek jet in loop quantum black~hole},'' \href{http://dx.doi.org/10.1088/1475-7516/2024/05/101}{{\em JCAP} {\bfseries 05} (2024) 101}, \href{http://arxiv.org/abs/2402.08402}{{\ttfamily arXiv:2402.08402 [astro-ph.HE]}}.

\bibitem{Zeng:2025olq}
X.-X. Zeng and K.~Wang, ``{Energy extraction from the Kerr-Bertotti-Robinson black hole via magnetic reconnection in a circular and a plunging plasma},'' \href{http://dx.doi.org/10.1103/vc96-snjm}{{\em Phys. Rev. D} {\bfseries 112} no.~6, (2025) 064032}, \href{http://arxiv.org/abs/2507.21777}{{\ttfamily arXiv:2507.21777 [gr-qc]}}.

\bibitem{Fan:2024rsa}
Z.-Y. Fan, F.~Zhou, Y.~Li, M.~Guo, and B.~Chen, ``{Magnetic reconnection under centrifugal and gravitational electromotive forces},'' \href{http://dx.doi.org/10.1103/PhysRevD.111.064067}{{\em Phys. Rev. D} {\bfseries 111} no.~6, (2025) 064067}, \href{http://arxiv.org/abs/2411.19491}{{\ttfamily arXiv:2411.19491 [astro-ph.HE]}}.

\bibitem{Zhang:2024rvk}
S.-J. Zhang, ``{Energy extraction via magnetic reconnection in Konoplya-Rezzolla-Zhidenko parametrized black holes},'' \href{http://dx.doi.org/10.1103/PhysRevD.109.084066}{{\em Phys. Rev. D} {\bfseries 109} no.~8, (2024) 084066}, \href{http://arxiv.org/abs/2402.15050}{{\ttfamily arXiv:2402.15050 [gr-qc]}}.

\bibitem{Camilloni:2024tny}
F.~Camilloni and L.~Rezzolla, ``{Self-consistent Multidimensional Penrose Process Driven by Magnetic Reconnection},'' \href{http://dx.doi.org/10.3847/2041-8213/adbbef}{{\em Astrophys. J. Lett.} {\bfseries 982} no.~1, (2025) L31}, \href{http://arxiv.org/abs/2411.04184}{{\ttfamily arXiv:2411.04184 [gr-qc]}}.

\bibitem{EventHorizonTelescope:2019dse}
{\bfseries Event Horizon Telescope} Collaboration, K.~Akiyama {\em et~al.}, ``{First M87 Event Horizon Telescope Results. I. The Shadow of the Supermassive Black Hole},'' \href{http://dx.doi.org/10.3847/2041-8213/ab0ec7}{{\em Astrophys. J. Lett.} {\bfseries 875} (2019) L1}, \href{http://arxiv.org/abs/1906.11238}{{\ttfamily arXiv:1906.11238 [astro-ph.GA]}}.

\bibitem{EventHorizonTelescope:2022wkp}
{\bfseries Event Horizon Telescope} Collaboration, K.~Akiyama {\em et~al.}, ``{First Sagittarius A* Event Horizon Telescope Results. I. The Shadow of the Supermassive Black Hole in the Center of the Milky Way},'' \href{http://dx.doi.org/10.3847/2041-8213/ac6674}{{\em Astrophys. J. Lett.} {\bfseries 930} no.~2, (2022) L12}, \href{http://arxiv.org/abs/2311.08680}{{\ttfamily arXiv:2311.08680 [astro-ph.HE]}}.

\bibitem{GRAVITY:2018ofz}
{\bfseries GRAVITY} Collaboration, R.~Abuter {\em et~al.}, ``{Detection of the gravitational redshift in the orbit of the star S2 near the Galactic centre massive black hole},'' \href{http://dx.doi.org/10.1051/0004-6361/201833718}{{\em Astron. Astrophys.} {\bfseries 615} (2018) L15}, \href{http://arxiv.org/abs/1807.09409}{{\ttfamily arXiv:1807.09409 [astro-ph.GA]}}.

\bibitem{abuter2018detection}
R.~Abuter, A.~Amorim, M.~Baub{\"o}ck, J.~Berger, H.~Bonnet, W.~Brandner, Y.~Cl{\'e}net, V.~C. Du~Foresto, P.~De~Zeeuw, C.~Deen, {\em et~al.}, ``Detection of orbital motions near the last stable circular orbit of the massive black hole sgra,'' {\em Astronomy \& Astrophysics} {\bfseries 618} (2018) L10.

\bibitem{GRAVITY:2020lpa}
{\bfseries GRAVITY} Collaboration, M.~Baub{\"o}ck {\em et~al.}, ``{Modeling the orbital motion of Sgr A*{\textquoteright}s near-infrared flares},'' \href{http://dx.doi.org/10.1051/0004-6361/201937233}{{\em Astron. Astrophys.} {\bfseries 635} (2020) A143}, \href{http://arxiv.org/abs/2002.08374}{{\ttfamily arXiv:2002.08374 [astro-ph.HE]}}.

\bibitem{Parfrey:2018dnc}
K.~Parfrey, A.~Philippov, and B.~Cerutti, ``{First-Principles Plasma Simulations of Black-Hole Jet Launching},'' \href{http://dx.doi.org/10.1103/PhysRevLett.122.035101}{{\em Phys. Rev. Lett.} {\bfseries 122} no.~3, (2019) 035101}, \href{http://arxiv.org/abs/1810.03613}{{\ttfamily arXiv:1810.03613 [astro-ph.HE]}}.

\bibitem{Komissarov:2005wj}
S.~S. Komissarov, ``{Observations of the Blandford-Znajek and the MHD Penrose processes in computer simulations of black hole magnetospheres},'' \href{http://dx.doi.org/10.1111/j.1365-2966.2005.08974.x}{{\em Mon. Not. Roy. Astron. Soc.} {\bfseries 359} (2005) 801--808}, \href{http://arxiv.org/abs/astro-ph/0501599}{{\ttfamily arXiv:astro-ph/0501599}}.

\bibitem{East:2018ayf}
W.~E. East and H.~Yang, ``{Magnetosphere of a spinning black hole and the role of the current sheet},'' \href{http://dx.doi.org/10.1103/PhysRevD.98.023008}{{\em Phys. Rev. D} {\bfseries 98} no.~2, (2018) 023008}, \href{http://arxiv.org/abs/1805.05952}{{\ttfamily arXiv:1805.05952 [astro-ph.HE]}}.

\bibitem{Hu:2020usx}
Z.~Hu, Z.~Zhong, P.-C. Li, M.~Guo, and B.~Chen, ``{QED effect on a black hole shadow},'' \href{http://dx.doi.org/10.1103/PhysRevD.103.044057}{{\em Phys. Rev. D} {\bfseries 103} no.~4, (2021) 044057}, \href{http://arxiv.org/abs/2012.07022}{{\ttfamily arXiv:2012.07022 [gr-qc]}}.

\bibitem{Hou:2022eev}
Y.~Hou, Z.~Zhang, H.~Yan, M.~Guo, and B.~Chen, ``{Image of a Kerr-Melvin black hole with a thin accretion disk},'' \href{http://dx.doi.org/10.1103/PhysRevD.106.064058}{{\em Phys. Rev. D} {\bfseries 106} no.~6, (2022) 064058}, \href{http://arxiv.org/abs/2206.13744}{{\ttfamily arXiv:2206.13744 [gr-qc]}}.

\bibitem{Huang:2024wpj}
J.~Huang, Z.~Zhang, M.~Guo, and B.~Chen, ``{Images and flares of geodesic hot spots around a Kerr black hole},'' \href{http://dx.doi.org/10.1103/PhysRevD.109.124062}{{\em Phys. Rev. D} {\bfseries 109} no.~12, (2024) 124062}, \href{http://arxiv.org/abs/2402.16293}{{\ttfamily arXiv:2402.16293 [gr-qc]}}.

\bibitem{Hu:2025hdr}
S.~Hu, D.~Li, and C.~Deng, ``{Light Curves of Chaotic Charged Hot-Spots in Curved Spacetime: Opening an Observational Window to Chaos},'' \href{http://arxiv.org/abs/2508.17384}{{\ttfamily arXiv:2508.17384 [gr-qc]}}.

\bibitem{Xie:2025skg}
F.~Xie, Q.-H. Zhu, and X.~Li, ``{Investigating non-Keplerian motion in flare events with astrometric data},'' \href{http://arxiv.org/abs/2507.07411}{{\ttfamily arXiv:2507.07411 [astro-ph.HE]}}.

\bibitem{Hu:2025lyp}
S.~Hu, S.~Tan, D.~Li, L.~Zhang, C.~Deng, and W.~Cao, ``{OCTOPUS: A Versatile, User-Friendly, and Extensible Public Code for General-Relativistic Ray-Tracing in Spherically Symmetric and Static Spacetimes},'' \href{http://arxiv.org/abs/2510.12585}{{\ttfamily arXiv:2510.12585 [gr-qc]}}.

\end{thebibliography}\endgroup

\end{document}